# A semi-Lagrangian implicit BGK collision model for the finite volume discrete Boltzmann method

Leitao Chen[1], Sauro Succi[2], Xiaofeng Cai[3], Laura Schaefer[1]

[1]*Department of Mechanical Engineering, Rice University, Houston, Texas 77005, USA*

[2]*Center for Life Nanoscience at La Sapienza, Italian Institute of Technology, 00161, Rome, Italy*

[3]*Department of Mathematical Science, University of Delaware, Newark, Delaware 19716, USA*

**Abstract**: In order to increase the accuracy of temporal solutions, reduce the computational cost of time marching and improve the stability associated to collisions for the finite volume discrete Boltzmann method (FVDBM), a new implicit BGK collision model using a semi-Lagrangian approach is proposed in this paper. Unlike existing models, in which the implicit BGK collision is resolved either by a temporal extrapolation or by a variable transformation, the new model removes the implicitness by tracing the particle distribution functions (PDFs) back in time along their characteristic paths during the collision process. An interpolation scheme is needed to evaluate the PDFs at the traced-back locations. By using the first-order interpolation, the resulting model allows for the straightforward replacement of $f_\alpha^{eq,n+1}$ by $f_\alpha^{eq,n}$ no matter where it appears. After comparing the new model with the existing models under different numerical conditions (e.g. different flux schemes and time marching schemes) and using the new model to successfully modify the variable transformation technique, three conclusions can be drawn. First, the new model can improve the accuracy by almost an order of magnitude. Second, it can slightly reduce the computational cost. Therefore, the new scheme improves accuracy without extra cost. Finally, the new model can significantly improve the Δt/τ limit compared to the temporal interpolation model while having the same Δt/τ limit as the variable transformation approach. The new scheme with a second-order interpolation is also developed and tested; however, that technique displays no advantage over the simple first-order interpolation approach. Both numerical and theoretical analyses are also provided to explain why the new implicit scheme with simple first-order interpolation can outperform the same scheme with second-order interpolation, as well as the existing temporal extrapolation and variable transformation schemes.

## 1. Background

Since its earliest development more than three decades ago [1-4], the lattice Boltzmann method (LBM) has gained a prominent role in the simulations of a large variety of complex flows across a broad range of scales, from macroscopic turbulence, all the way down to nanoscale flows of biological interest, and lately, even sub-nuclear flows [5]. Its success is supported by two important features. First, physically, the LBM can inherently solve problems over a wide range of length scales beyond the strict hydrodynamic regime [6]. The behavior of hydrodynamics at macroscales is basically a low-dimensional asymptotic limit of the infinite-dimensional sequence of kinetic moments associated with the Boltzmann equation that is rooted in the microscale kinetics. By capturing the high-order moments in the Boltzmann equation, the low-order moments in macroscale hydrodynamics emerge naturally from the underlying micro-dynamics [7]. This is why the LBM is regarded as a mesoscale technique with both upwards (to the continuum) and downwards (to the atomistic) multiscale capabilities. Second, numerically, the LBM can achieve second-order accuracy in space with only a first-order numerical scheme [8]. The reason for this is that the advection term $\boldsymbol{e}_\alpha \cdot \boldsymbol{\nabla} f_\alpha$ in the LBM is linear ($\boldsymbol{e}_\alpha$ before the gradient is constant). Due to the linear advection, the LBM can couple the discretizations of all three dimensions: the microscopic velocity ($\boldsymbol{e}$), space ($\boldsymbol{x}$), and time ($t$). By doing this, the variables that are being advected (in this case, the particle distribution functions) will stop exactly at a grid point after each advection step. According to the definition of the Courant-Friedrichs-Lewy (*CFL*) number, the *CFL* of the microscopic velocities in the LBM becomes one globally, which gives rise to a universal second-order accuracy in space.

Although this unique multi-dimensional coupling mechanism is an important asset of the LBM, it also brings with it a substantial challenge. Since the LBM couples the discretizations of all three dimensions, this limits the freedom to choose a different way of individually discretizing any of the three dimensions, which is especially restrictive for the spatial dimension $\boldsymbol{x}$. Therefore, the mesh, which is the result of discretizing the space, has to copy the lattice structure (a lattice tells how the velocity is discretized), and also has to be uniform (in order to achieve *CFL*=1 location-wise) and rigid (in order to achieve *CFL*=1 time-wise). Consequently, such a uniform and rigid mesh structure makes it difficult for the LBM to accurately accommodate problems with curved or complicated boundaries [9], which are, however, ubiquitous in fluid flow problems.

Numerically speaking, the LBM (with its coupling feature mentioned above) is derived from the discrete Boltzmann equation (DBE) whose space and time are still continuous. Therefore, by solving the DBE, one can select an arbitrary discretization for space. As a result, complex boundaries can be easily captured with a body-fitted mesh, just like in the conventional computational fluid dynamics (CFD) models. The earliest work following this idea was presented by Nannelli and Succi [10], in which the finite volume method (FVM) is applied on irregular meshes. Their work belongs to a category called the finite volume discrete Boltzmann method (FVDBM), which has witnessed a rapid progression in the following decades [11-21]. Another major method that could also use this approach is the finite element discrete Boltzmann method (FEDBM) [22-24] because the finite element method (FEM) can easily integrate unstructured meshes as well. However, this has not gained the same popularity as the FVDBM due to the mathematical simplicity and the built-in conservation of the FVM.

Unfortunately, as a result of the mesh flexibility, the FVDBM (as well as other methods based on solving the DBE) currently exhibits a lower accuracy and higher computational cost than the LBM. Since space and time in the FVDBM are decoupled, the accuracy and computational cost in space as well as in time must be handled separately.

In the space dimension, the accuracy of the FVDBM is mostly limited by the diffusion error. Per a Fourier stability analysis, this diffusion error automatically appears when *CFL*<1, which is required to maintain proper stability when solving the DBE on irregular meshes [25]. Such a diffusion error in the FVDBM has been well acknowledged since the early stages of its development [15, 26]. However, very few publications have provided solutions on how to reduce this diffusion error at a reasonable cost. As a result, it has been asserted that the FVDBM is not a competitive alternative to the LBM [17]. As an effort to address this issue, one of our previous papers provided a systematic approach that could produce Godunov-type flux schemes with different orders of accuracy for the advection in the FVDBM, which could significantly reduce the diffusion error beyond that of the conventional upwinding schemes [27]. We also developed a new second-order interpolation scheme designated the plane-fitting least-square (PFLS) approach to reduce the diffusion error during the interpolation step of the FVDBM [28], which displayed a faster speed as well as a slightly better accuracy than the conventional least-square interpolation scheme.

In the time dimension, the time marching scheme should be carefully chosen, since when solving the DBE, the maximum $\Delta t$ is not only limited by the *CFL*, which is controlled by the advection, but also limited by the relaxation time, which is affected by the collision. The explanation is that $\Delta t$, which is the numerical time interval for updating the solution, cannot be too large compared to the relaxation time, which is the physical time that the system takes during each time step to relax towards the equilibrium state. As a result, solving the DBE requires a very small $\Delta t$ when modeling steady-state high-Re flows in which the relaxation time is very small. Therefore, the selected time marching scheme should allow the use of a $\Delta t$ that is as large as possible, as long it is within the physical limit. The standard approach to achieve this is to make the collision implicit for the time marching [22, 23, 30], which, however, creates an implicit (nonlinear) equilibrium term that requires additional treatment. Currently, there are basically two approaches to resolve this implicitness: the temporal extrapolation (TE) scheme that calculates the implicit value based on two previous time steps [30], and the variable transformation (VT) technique that can wrap the implicit term into a new variable [31-40]. It was reported that the VT can dramatically improve the stability beyond the TE scheme [35, 40]. However, the comparison in accuracy and computational cost between these two is unknown.

In this paper, we develop a new scheme to resolve the implicit collision during time marching. The new scheme is based on applying the semi-Lagrangian (SL) treatment to the implicit collision term. Therefore, it is designated as the semi-Lagrangian implicit collision (SLIC) model in this paper. After a quantitative comparison in accuracy, computational cost, and stability between the new scheme and existing approaches on the Bhatnagar-Gross-Krook (BGK) collision model [29], it is found that the new scheme is more accurate and slightly faster than the existing schemes. It is also shown that the new scheme is more stable than the TE scheme while having the same stability as the VT scheme.

## 2. The FVDBM with an implicit BGK collision

The DBE with the BGK collision model is defined as:

$$\frac{\partial f_\alpha}{\partial t} + \boldsymbol{e}_\alpha \cdot \boldsymbol{\nabla} f_\alpha = -\frac{1}{\tau}\left(f_\alpha - f_\alpha^{eq}\right) \qquad \alpha = 0,1,2,\dots,N-1 \qquad (1)$$

where $f_\alpha$ and $f_\alpha^{eq}$ are the particle distribution function (PDF) and equilibrium PDF, respectively, in the $\alpha$th direction of a total of $N$ components, $\boldsymbol{e}_\alpha$ is the $\alpha$th of $N$ total lattice velocities, and $\tau$ is the relaxation time.

With the help of the FVM, Eq. (1) can be integrated over a control volume (CV). Then, after a rearrangement, the FVDBM in its general form is shown as:

$$T_\alpha = C_\alpha - F_\alpha \tag{2}$$

where $T_\alpha$, $C_\alpha$, and $F_\alpha$ are the temporal term, collision term, and flux term, respectively. The temporal and collision terms are:

$$T_\alpha = \frac{\partial f_\alpha}{\partial t} \tag{3}$$

$$C_\alpha = \frac{1}{\tau}(f_\alpha^{eq} - f_\alpha) \tag{4}$$

It is worth noting that, so far, Eq. (2) is still continuous both in space and time. When discretizing the space with a mesh such as an unstructured one, the total flux through the surface closure of each CV becomes the summation of the flux through each of the total $K$ surface segments of the CV. Then the flux term in Eq. (2) becomes:

$$F_\alpha = \frac{1}{V_{CV}} \sum_{i=1}^{K} F_{\alpha,i} \tag{5}$$

where $V_{CV}$ is the volume of the CV. In the current study, cell-centered triangular meshes are used, which makes $K = 3$. After discretizing the time with a proper time marching scheme, Eq. (2) can be solved numerically. With the standard forward Euler method, Eq. (2) becomes:

$$T_\alpha^n = C_\alpha^n - F_\alpha^n \tag{6}$$

where

$$T_\alpha^n = \frac{f_\alpha^{n+1} - f_\alpha^n}{\Delta t} \tag{7}$$

and $C_\alpha^n$ and $F_\alpha^n$ are the collision and flux terms that are evaluated at time step $t_n$. By inserting Eq. (7) into Eq. (6), replacing $C_\alpha^n$ with its definition Eq. (4), and combining the terms that contain $f_\alpha^n$, the simplest form for the FVDBM is obtained:

$$f_\alpha^{n+1} = \left(1 - \frac{\Delta t}{\tau}\right) f_\alpha^n + \frac{\Delta t}{\tau} f_\alpha^{eq,n} - \Delta t F_\alpha^n \tag{8}$$

However, it is well known that the forward Euler method is explicit and is not as stable as implicit methods. Lee et al. [22, 23], Guo et al. [35], and Bardow et al. [31] tried to introduce implicitness into the system. They developed a general formula that keeps the collision implicit and the advection explicit. After applying this method to the FVDBM, it becomes:

$$T_\alpha^n = [(1 - \theta)C_\alpha^n + \theta C_\alpha^{n+1}] - F_\alpha^n \tag{9}$$

where $\theta$ is a tuning parameter that varies between 0 and 1. The collision term in Eq. (9) becomes fully explicit if $\theta = 0$, and fully implicit once $\theta = 1$. Here we start from a simple case in which $\theta = 1$. Then Eq. (9) becomes:

$$T_\alpha^n = C_\alpha^{n+1} - F_\alpha^n \tag{10}$$

By combining Eqs. (4) and (7) into Eq. (10) to replace the collision and temporal terms and rearranging the equation, Eq. (10) becomes:

$$f_\alpha^{n+1} = \frac{\tau}{\tau+\Delta t} f_\alpha^n + \frac{\Delta t}{\tau+\Delta t} f_\alpha^{eq,n+1} - \frac{\tau \Delta t}{\tau+\Delta t} F_\alpha^n \tag{11}$$

Eq. (11) is more stable than Eq. (8). However, there is still implicitness left untouched in $f_\alpha^{eq,n+1}$ that needs to be resolved, which will be discussed in the next two sections. It is important to note that a proper flux scheme is required to calculate the flux term $F_\alpha^n$ in order to close the system, which will also be discussed later.

## 3. The state of art of resolving the implicitness in $f_\alpha^{eq,n+1}$

The most simple and straightforward approach to resolve implicitness for any problem is to solve the implicit variables with an iterative process. For the current application, the procedure should be performed with the following steps:

Step 1: Guess an initial value for $f_\alpha^{eq,n+1}$;
Step 2: Calculate $f_\alpha^{n+1}$ with Eq. (11);
Step 3: Calculate the moments with $f_\alpha^{n+1}$ from step 2;
Step 4: Calculate the new $f_\alpha^{eq,n+1}$ with the moments from step 3;
Step 5: Check the difference between the new $f_\alpha^{eq,n+1}$ and its value in the last iteration. If it is converged, finish; otherwise, repeat steps 2 to 5.

This iterative process is very costly since the calculation of moments (step 3) and the calculation of the equilibrium PDF (step 4) are computationally intense and the convergence criteria must be met at all grid locations. Therefore, this method is not studied in this paper. Instead, all the methods discussed in this paper are non-iterative.

### 3.1 The temporal extrapolation (TE) scheme

This approach directly solves $f_\alpha^{eq,n+1}$ and then substitutes it back into Eq. (11) to close the system. According to Mei and Shyy [30], $f_\alpha^{eq,n+1}$ can be linearly extrapolated, as a whole, by using its own values in the two previous time steps, namely:

$$f_\alpha^{eq,n+1} = 2f_\alpha^{eq,n} - f_\alpha^{eq,n-1} \tag{12}$$

which is termed the temporal extrapolation (TE) scheme in this paper. For the standard two-dimensional nine-velocity model (D2Q9), the equilibrium PDF at any time step is computed as:

$$f_\alpha^{eq} = \omega_\alpha \rho \left[ 1 + \frac{\boldsymbol{e}_\alpha \cdot \boldsymbol{u}}{c_s^2} + \frac{(\boldsymbol{e}_\alpha \cdot \boldsymbol{u})^2}{2c_s^4} - \frac{\boldsymbol{u} \cdot \boldsymbol{u}}{2c_s^2} \right] \tag{13}$$

where $\omega_\alpha$ is the weight in each corresponding direction, $c_s$ is the speed of sound, and $\boldsymbol{u}$ and $\rho$ are the macroscopic velocity and density, or moments, which can be calculated as:

$$\begin{bmatrix} \rho \\ \rho\boldsymbol{u} \end{bmatrix} = \Sigma_{\alpha=0}^{N-1} \begin{bmatrix} f_\alpha \\ \boldsymbol{e}_\alpha f_\alpha \end{bmatrix} \tag{14}$$

As a result, the computation procedure of the FVDBM using the TE scheme during each time step is as follows:

Step 1: Calculate the moments with Eq. (14) with the newest $f_\alpha$ ;
Step 2: Calculate $f_\alpha^{eq}$ with Eq. (13) based on the moments from step 1;
Step 3: Calculate $f_\alpha^{eq,n+1}$ with Eq. (12);
Step 4: Update $f_\alpha$ with Eq. (11) with the $f_\alpha^{eq,n+1}$ from step 3.

It should be noted that step 2 requires only one computation of $f_\alpha^{eq}$ but additional memory allocation to store its value at $t_{n-1}$.

**3.2 The variable transformation (VT) scheme**

The TE scheme is very easy to implement. However, it was noted by Mei and Shyy [30] that the TE scheme is prone to instability due to the extrapolation. In order to address this, He *et al.* [32] introduced a technique called variable transformation (VT) to avoid the need for a temporal extrapolation. It has been shown that the VT scheme is much more stable than the TE approach [35, 40], and therefore has become a widely accepted method [31-40]. The VT scheme does not focus on solving $f_\alpha^{eq,n+1}$ by itself. Instead, it treats the entire collision term as a whole. In the context of the FVDBM, the VT scheme starts from the following governing equation, which is the result of replacing $T_\alpha^n$ in Eq. (10) with Eq. (7):

$$f_\alpha^{n+1} = f_\alpha^n + \Delta t C_\alpha^{n+1} - \Delta t F_\alpha^n \tag{15}$$

By defining a new variable $g_\alpha$ that contains the collision as:

$$g_\alpha = f_\alpha - \Delta t C_\alpha \tag{16}$$

at the time step $t_{n+1}$, it holds that:

$$g_\alpha^{n+1} = f_\alpha^{n+1} - \Delta t C_\alpha^{n+1} \tag{17}$$

and by combining Eq. (17) and Eq. (15), this becomes:

$$g_\alpha^{n+1} = f_\alpha^n - \Delta t F_\alpha^n \tag{18}$$

It can be seen that there is no implicitness on the right-hand side (RHS) of Eq. (18), so $g_\alpha^{n+1}$ can be computed after the flux calculation is finished. The next task is to recover $f_\alpha^{n+1}$ from $g_\alpha^{n+1}$. By rearranging Eq. (17) after expanding the BGK collision term and combining the terms that contain $f_\alpha^{n+1}$, it can be obtained that:

$$f_\alpha^{n+1} = \frac{\tau}{\tau+\Delta t}\left(g_\alpha^{n+1} + \frac{\Delta t}{\tau} f_\alpha^{eq,n+1}\right) \quad (19)$$

The new variable $g_\alpha$ satisfies the condition that it preserves the moments of $f_\alpha$, therefore:

$$\begin{bmatrix}\rho \\ \rho\boldsymbol{u}\end{bmatrix} = \sum_{\alpha=0}^{N-1}\begin{bmatrix}g_\alpha \\ \boldsymbol{e}_\alpha g_\alpha\end{bmatrix} = \sum_{\alpha=0}^{N-1}\begin{bmatrix}f_\alpha \\ \boldsymbol{e}_\alpha f_\alpha\end{bmatrix} \quad (20)$$

Since the equilibrium PDF can be exclusively determined by its moments, as shown in Eq. (13), it also holds that:

$$f_\alpha^{eq,n+1} = g_\alpha^{eq,n+1} \quad (21)$$

As a result, the procedure of using the VT scheme for the FVDBM during each time step is:

Step 1: Calculate $g_\alpha^{n+1}$ with Eq. (18);
Step 2: Calculate the moments based on $g_\alpha^{n+1}$ with Eq. (20) or Eq. (14);
Step 3: Calculate $g_\alpha^{eq,n+1}$ with Eq. (13) based on the moments from step 2;
Step 4: Update $f_\alpha$ with Eq. (19) by applying Eq. (21).

**4. The semi-Lagrangian implicit collision (SLIC) scheme**

The semi-Lagrangian (SL) method originated in the applied math community for solving the general transport equation [41-44]. It preserves the mesh flexibility of the Eulerian method while maintaining a good level of accuracy and large *CFL* numbers of the Lagrangian method. And this why it is "semi". The earliest applications of the semi-Lagrangian method in the LBM community were introduced by Shu *et al*. [45] and Cheng *et al*. [46], in which the interpolation was introduced during the streaming step in order to remove the restriction imposed by the rigid mesh structure. Recently Krämer *et al*. [47] also applied the SL method to the streaming on their off-lattice Boltzmann method (OLBM) and found that the SL method can increase the computational efficiency by allowing a larger time step size. In 2018, Di Ilio *et al*. [48] chose the SL method for streaming to study turbulent flows with a body-fitted mesh for complex geometries. In the same year, Dorschner *et al.* [49] applied the SL method to the advection in their "particles on demand" framework in order to remove the limitation of flow speed and temperature range in the original LBM, which is the most recent work on the applications of SL method in the LBM as of the writing of this paper. However, all existing applications of the SL method are only for advection, and it has never been applied to the collision of any LBM or DBM work. In this section, a new approach that resolves the implicitness by applying the SL method to the implicit collision term is developed. This approach is called the semi-Lagrangian implicit collision (SLIC) in this paper and completely different from the TE and VT methods discussed in the previous section. The development of the SLIC scheme will be explained in detail in the rest of this section by starting from the re-examination of $f^{eq}$, which is defined as the Maxwellian distribution that is a function of moments (density $\rho$, macroscopic velocity $\boldsymbol{u}$, etc.) such that:

$$f^{eq} = \frac{\rho}{(2\pi RT)^{\frac{D}{2}}} exp\left[-\frac{(\boldsymbol{e}-\boldsymbol{u})^2}{2RT}\right] \quad (22)$$

where $T$ is the temperature, $D$ is the degree of spatial dimensions, and $R$ is the ideal gas constant. The microscopic velocity $e$ in Eq. (22) is still continuous, which needs to be discretized in order to be solved computationally. Once discretized, $f^{eq}$ can be computed by performing a Taylor expansion on the Maxwellian. Eq. (13) is the discrete form of $f^{eq}$ on the D2Q9 lattice with a second-order truncation. After the discretization, the moments can be recovered by taking an ensemble of the PDFs, as shown in Eq. (14).

From Eqs. (13) and (14) it can be seen that $f_\alpha^{eq}$ is a function of moments that is further a function of $f_\alpha$, which can be depicted by the following notation:

$$f_\alpha^{eq} = M\{m[f_\alpha]\} \tag{23}$$

where $m[\ ]$ is the operator that calculates the moments from the PDFs, and $M\{\ \}$ is the operator that calculates the equilibrium PDFs from the moments. (Technically speaking, $M\{\ \}$ is the Maxwellian operator that calculates the equilibrium PDFs by Eq. (22). However, its notation is borrowed here to specifically represent the calculation of equilibrium PDFs with discrete velocities). As shown in Eq. (23), $f_\alpha^{eq}$ is an indirect function of $f_\alpha$, which is at the same time the solution of the DBE (Eq. (1)). This is the reason why the implicitness in $f_\alpha^{eq,n+1}$ is difficult to treat.

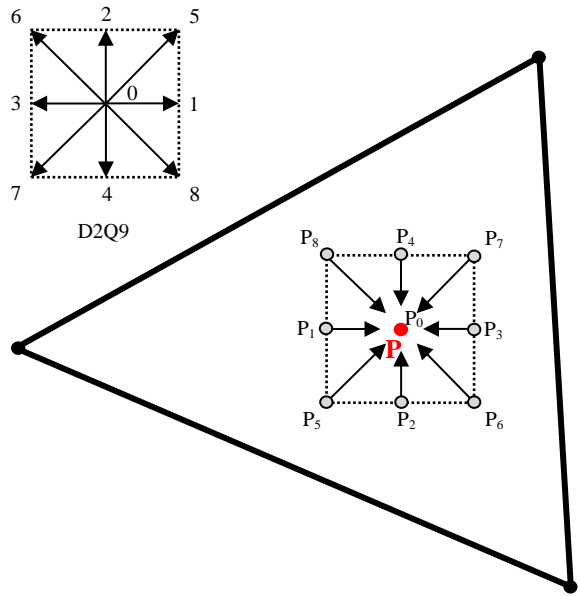

*Figure 1. Advection of PDFs along characteristics*

Unlike the TE and VT schemes, the SLIC method tracks the PDFs along their characteristics back in time in a Lagrangian way and was initially proposed by Groppi *et al.* [50]. The SLIC method consists of two steps, which will be explained as follows. First, according to Eq. (23), at the barycenter $P$ of any CV, as shown in Fig. 1, and at $t_{n+1}$, it holds that:

$$f_\alpha^{eq,n+1}(P) = M\{m[f_\alpha^{n+1}(P)]\} \tag{24}$$

Second, as pointed out by Groppi *et al.* [50], the PDFs preserve their values along their characteristic lines when being advected or streaming, followed by their moments. Therefore, the PDFs at $t_{n+1}$ are the same as the PDFs at $t_n$ at their previous locations rendered by being tracked back in time along the characteristic paths. Taking the D2Q9 lattice model as an example, whose structure and numbering of different directions are shown in the inset of Fig. 1, the nine PDFs that rendezvous at the location $P$ at $t_{n+1}$ were advected from different locations ($P_0$ to $P_8$) at the previous time step $t_n$. Therefore, it holds that:

$$f_\alpha^{n+1}(P) = f_\alpha^n(P_\alpha) \tag{25}$$

Since the PDFs keep their values along their characteristic paths, the moments that are calculated based on these PDFs also stay the same. Therefore, there exists:

$$m[f_\alpha^{n+1}(P)] = m[f_\alpha^n(P_\alpha)] \tag{26}$$

For the D2Q9 lattice, Eq. (26) means that:

$$\begin{bmatrix} \rho^{n+1}(P) \\ \rho^{n+1}(P)\boldsymbol{u}^{n+1}(P) \end{bmatrix} = \sum_{\alpha=0}^{8} \begin{bmatrix} f_\alpha^n(P_\alpha) \\ \boldsymbol{e}_\alpha f_\alpha^n(P_\alpha) \end{bmatrix} \tag{27}$$

The tracked-back locations, $P_0$ to $P_8$ ($P_0$ is the same location as $P$ since $\boldsymbol{e}_0$ is **0**), are generally not located at grid points (in the cell-centered meshes, they are not located at barycenters). Therefore, an interpolation scheme is needed to evaluate $f_\alpha^n(P_\alpha)$. If using $X$ to denote the coordinate of a point, then the coordinates of $P_\alpha$ are known as:

$$\boldsymbol{X}(P_\alpha) = \boldsymbol{X}(P) - \boldsymbol{e}_\alpha \Delta t \tag{28}$$

This location information is deterministic and can be used to calculate $f_\alpha^n(P_\alpha)$ with a chosen interpolation scheme. Once $f_\alpha^n(P_\alpha)$ becomes known, the implicitness can be closed as:

$$f_\alpha^{eq,n+1}(P) = M\{m[f_\alpha^n(P_\alpha)]\} \tag{29}$$

If a first-order interpolation scheme is selected, which means the PDF distributions are constant within each CV, it can be assumed that:

$$f_\alpha^n(P_\alpha) = f_\alpha^n(P) \tag{30}$$

which says the PDF at the tracked-back locations within the same CV is equal to its value at the barycenter of the CV. Finally, Eq. (29) can be reduced to:

$$f_\alpha^{eq,n+1}(P) = M\{m[f_\alpha^n(P)]\} \tag{31}$$

By revisiting Eq. (23) for the definition of $f_\alpha^{eq}$, it can be seen that the RHS of Eq. (31) is actually $f_\alpha^{eq,n}(P)$. Therefore, Eq. (31) can be further reduced to:

$$f_\alpha^{eq,n+1}(P) = f_\alpha^{eq,n}(P) \quad or \quad f_\alpha^{eq,n+1} = f_\alpha^{eq,n} \tag{32}$$

Eq. (32) is the final form of the SLIC scheme with the first-order interpolation. For convenience, this is designated as SLIC+INT1 in this paper. The numerical sequence of updating the FVDBM solution with the SLIC+INT1 scheme during each time is:

Step 1: Calculate the moments with Eq. (14) with the newest $f_\alpha$ ;
Step 2: Calculate $f_\alpha^{eq,n}$ with Eq. (13) based on the moments from step 1;
Step 3: Update $f_\alpha$ with Eq. (11) by using Eq. (32) for $f_\alpha^{eq,n+1}$.

One can also choose a second-order interpolation to close the SLIC scheme. Here we propose to use the PFLS scheme from [28], in which the PDF distribution is assumed to be a linear function such as:

$$f_\alpha(x, y) = c_1 + c_2 \frac{x-x_0}{\sqrt{\Delta_0}} + c_3 \frac{y-y_0}{\sqrt{\Delta_0}} \tag{33}$$

where $(x, y)$ and $(x_0, y_0)$ are the coordinates of any tracked-back location $X$ and the barycenter $P$, respectively, as shown in Fig. 2, and $\Delta_0$ is the area of the center CV whose barycenter is $P$. In order to calculate the PDFs at the tracked-back locations, the coefficients $c_1$, $c_2$, and $c_3$ in Eq. (33) need to be determined. With the help of three neighbor CVs whose barycenters are $N_1(x_1, y_1)$, $N_2(x_2, y_2)$, and $N_3(x_3, y_3)$, these coefficients can be calculated as:

$$[c_1 \quad c_2 \quad c_3]' = (\boldsymbol{G}^T\boldsymbol{G})^{-1}\boldsymbol{G}^T\boldsymbol{F} \tag{34}$$

where

$$\boldsymbol{G} = \begin{pmatrix} 1 & 0 & 0 \\ 1 & \frac{x_1-x_0}{\sqrt{\Delta_0}} & \frac{y_1-y_0}{\sqrt{\Delta_0}} \\ 1 & \frac{x_2-x_0}{\sqrt{\Delta_0}} & \frac{y_2-y_0}{\sqrt{\Delta_0}} \\ 1 & \frac{x_3-x_0}{\sqrt{\Delta_0}} & \frac{y_3-y_0}{\sqrt{\Delta_0}} \end{pmatrix} \tag{35}$$

$$\boldsymbol{F} = \begin{pmatrix} f_\alpha(P) \\ f_\alpha(N_1) \\ f_\alpha(N_2) \\ f_\alpha(N_3) \end{pmatrix} \tag{36}$$

Once $c_1$, $c_2$, and $c_3$ are determined, the $f_\alpha^n(P_\alpha)$ in Eq. (29) can be calculated by plugging the coordinates of $P_\alpha$ into Eq. (33). This is the SLIC scheme with second-order interpolation, which will be designated SLIC+INT2 in this paper. It is worth noting that the matrix $\boldsymbol{G}$ (Eq. (35)) contains only geometric information. Therefore, the computation process can be optimized by pre-computing the entire grouping of $(\boldsymbol{G}^T\boldsymbol{G})^{-1}\boldsymbol{G}^T$ in Eq. (34) and storing it in memory. However, even with this optimization, the SLIC+INT2 scheme undoubtedly will be much slower than the SLIC+INT1 approach, since the former requires the computation of Eqs. (33) and (34) for multiple times (9 times for D2Q9) as well as Eq. (29) for one time, but the latter only requires the computation of $f_\alpha^{eq,n}$ by Eq. (13) for one time.

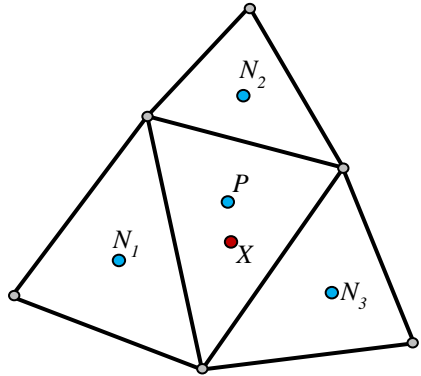

*Figure 2. Second-order interpolation for the tracked-back locations*

The procedure of updating the FVDBM solution with the SLIC+INT2 scheme during each time is:

Step 1: Calculate the coefficients with Eq. (34);
Step 2: Calculate the PDF at the tracked-back location with Eq. (33);
Step 3: Repeat step 1 and step 2 for all tracked-back locations;
Step 4: Gather the PDFs at all tracked-backed locations and combine them in Eq. (29) to compute $f_\alpha^{eq,n+1}$;
Step 5: Update $f_\alpha$ with Eq. (11) with the $f_\alpha^{eq,n+1}$ from step 4.

## 5. Simulation results and discussions: a preliminary study

Taylor-Green vortex (TGV) flow is chosen as the major example case for this study. The analytical velocities $(u_x, u_y)$ at any location $(x, y)$ and any time $t$ are defined as:

$$u_x = -u_0 \cos(k_1 x) \sin(k_2 y) \, e^{[-\nu(k_1^2+k_2^2)t]} \tag{37}$$

$$u_y = u_0 \frac{k_1}{k_2} \sin(k_1 x) \cos(k_2 y) \, e^{[-\nu(k_1^2+k_2^2)t]} \tag{38}$$

where $u_0$ is a reference velocity, $\nu$ is the kinematic viscosity, and $k_1$ and $k_2$ are defined as:

$$k_1 = \frac{2\pi}{D_x}, \qquad k_2 = \frac{2\pi}{D_y} \tag{39}$$

where $D_x$ and $D_y$ are the length and height of the flow domain. The SLIC+INT1 and SLIC+INT2 schemes are compared with the TE and VT schemes by solving the FVDBM with Eq. (11) ($\theta = 1$) for the TGV flow. The second-order upwind (SOU) is used for all schemes to calculate the flux term $F_\alpha^n$ in Eq. (11). The comparisons in terms of accuracy (temporal accuracy), computational cost, and stability are made for all schemes, which will be discussed in detail in the following subsections. Before performing the comparisons, an appropriate mesh size should be chosen. Therefore, a grid convergence study is undertaken for all implicit schemes as shown in Fig. 3. Except for the mesh resolution, all of the parameters are kept the same, including $\tau = 0.009$ and $\frac{\Delta t}{\tau} = 0.2$, and all of the errors are measured when $t = 0.5 t_c$ ($t_c$ is the time when the TGV flow has decayed to exactly 50% of its initial strength). By fixing $\tau$, the *Re* of the flow is also fixed due to the linear relation between $\tau$ and $\nu$, defined as $\nu = \tau c_s^2$. All of the curves in Fig. 3 can be described by the simple power law $ax^b$. After $a$ and $b$ for each curve are determined through curve fitting with a high confidence level of $R^2 = 0.99$, the pre-factor $\log(a)$ and slope $b$ of each curve can be obtained, as shown in Fig. 3. It can be seen that the SLIC+INT1 presents the steepest slope, which is -0.71. The mesh size chosen for all numerical studies in this paper is 40,000, since at this mesh resolution, the errors for all of the schemes are below a reasonable amount, which is $10^{-1}$. All numerical parameters used in the convergence study along with the selected mesh size are also applied in other studies in this paper except when otherwise noted. There are another two observations one can make in Fig. 3. First, the TE and VT schemes are almost identical in terms of error; and second, the SLIC+INT2 approach produces larger errors than SLIC+INT1. The analysis behind these observations will be given in subsequent subsections.

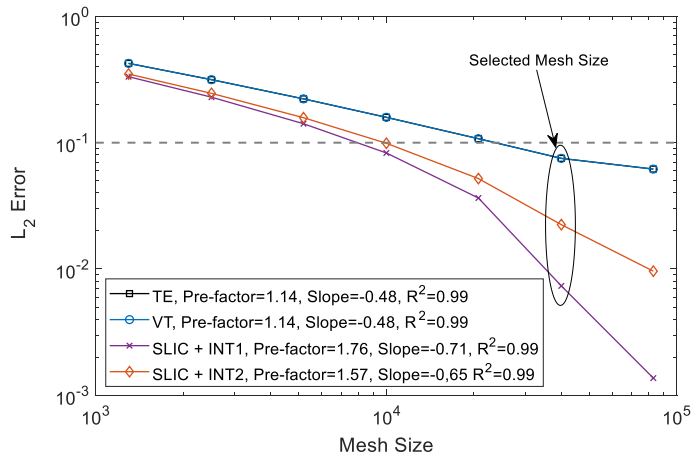

*Figure 3. Grid convergence study on the Taylor-Green vortex flow*

### 5.1 Accuracy

The L$_2$ errors (with respect to the analytical solution in Eqs. (37) and (38)) of the FVDBM transient solutions with four implicit collision schemes are calculated. In order to see the effect of $\Delta t$ on the transient solutions, the transient errors in the window from 0 to $0.5 t_c$ with four different sizes of $\Delta t$ are measured and compared in Fig. 4. It can be seen that at all sizes of $\Delta t$, the errors for all schemes grow with time because the transient errors will accumulate during each time step. However, the errors from the two SLIC schemes (especially the SLIC+INT1) grow at a slower pace than the TE and VT schemes as time progresses, which clearly indicates that the SLIC scheme can generate much less temporal error than the TE and VT schemes. At $\Delta t = 0.2\tau$, the error of the TE (or VT) scheme is more than eight times the error from the SLIC+INT1 scheme, which means the SLIC+INT1 can improve the temporal accuracy by a factor of eight. When taking an average among all of the plots in Fig. 4, this factor is almost four, which is still very high. As pointed out in the grid convergence study, the TE and VT schemes generate almost identical results, which can also be well observed in Fig. 4. In order to quantify their difference, the error difference between the TE and VT schemes (TE minus VT) of the transient solutions is plotted in Fig. 5, from which it can be seen that the VT scheme always generates slightly less error than the TE scheme, and that such a difference also grow with time and becomes larger with the increase of $\Delta t$.

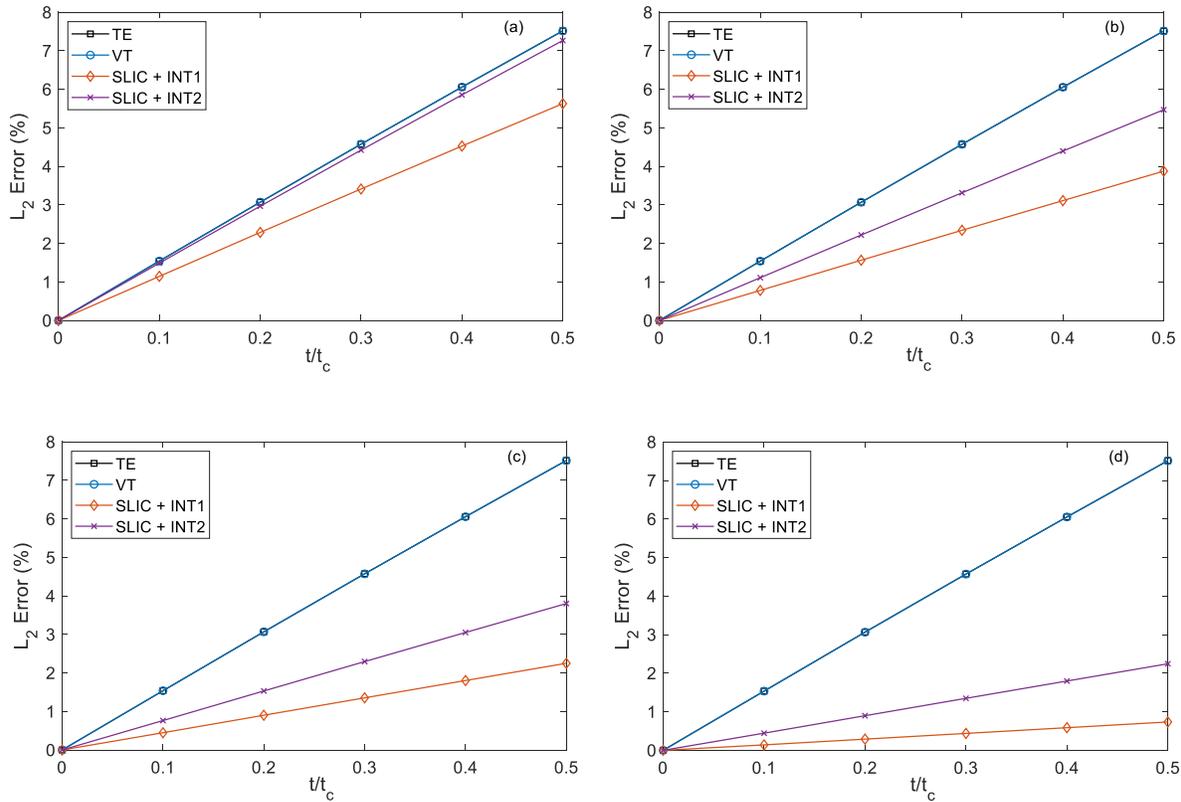

Figure 4. The $L_2$ errors of transient FVDBM solutions with different implicit collision schemes during the time span from 0 to $0.5t_c$ for (a) $\Delta t = 0.05\tau$; (b) $\Delta t = 0.1\tau$; (c) $\Delta t = 0.15\tau$; and (d) $\Delta t = 0.2\tau$

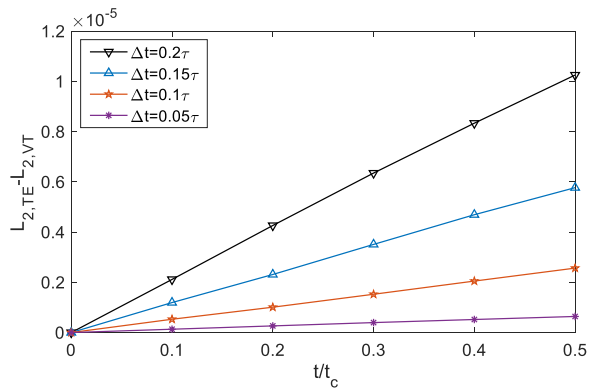

Figure 5. The $L_2$ error difference between the TE and VT schemes (TE-VT) of transient FVDBM solutions during the time span from 0 to $0.5t_c$

By examining each plot in Fig. 4 again, it is not difficult to conclude that the change of $\Delta t$ has different effects on the FVDBM solvers with different implicit collision schemes. For the TE and VT schemes, the change of $\Delta t$ barely affects the error. On the other hand, an increasing $\Delta t$ will decrease the error of the solver with the two SLIC schemes. These can be seen more clearly in Fig. 6, in which the errors with different schemes at $t = 0.5t_c$ are plotted against $\Delta t$. The error of a transient solution at any instance $t$ is the error accumulation from the initial time step to the current time step. In other words:

$$E_t = \sum_1^S \epsilon^n \tag{40}$$

where $E_t$ is the error of the transient solution measured at the current time, $\epsilon^n$ is the error generated during the $n$th time step, and $S$ is the total number of time steps from the initial to the current time. For the TE and VT schemes, although $\epsilon^n$ will increase with increasing $\Delta t$, it takes fewer time steps to reach the same end time $t$. As a result, the total accumulated error stays the same. Therefore, the FVDBM with the TE and VT schemes belongs to the family that satisfies the condition that:

$$\frac{\epsilon_1}{\epsilon_2} = \frac{\Delta t_1}{\Delta t_2} \tag{41}$$

where $\Delta t_1$ and $\Delta t_2$ are two different time step sizes, and $\epsilon_1$ and $\epsilon_2$ are the corresponding errors generated during one time step. On the contrary, the errors of the two SLIC schemes at $t = 0.5t_c$ dramatically decrease with an increase in $\Delta t$. At $\Delta t = 0.05\tau$, the SLIC+INT1 generates an error of less than 6% compared to the analytical solution, which is a roughly 25% decrease in error compared to the TE or VT scheme. When increasing the time step size to $\Delta t = 0.2\tau$, the SLIC+INT1 scheme produces an error of only about 1% compared to the analytical solution, which is an 85% cut in error compared to the TE and VT schemes. The SLIC schemes behave like this because they belong to another family of schemes that satisfies:

$$\frac{\epsilon_1}{\epsilon_2} > \frac{\Delta t_1}{\Delta t_2} \tag{42}$$

which means when the time step size increases the temporal error during each time step also increases but at a lower rate. This is an important feature since a larger $\Delta t$ will not only improve the accuracy of the solution but also bring down the computational cost by taking fewer time steps if a steady-state solution is sought. This feature of the SLIC scheme was also observed in the publication of Qiu & Shu [43] although the semi-Lagrangian method was applied to the advection term, not the collision.

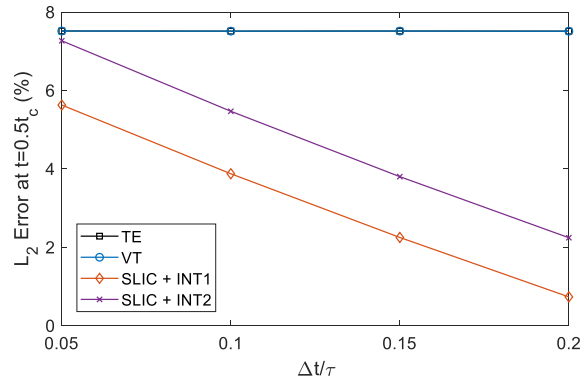

*Figure 6. The effects of $\Delta t$ on the error of the FVDBM solutions with different implicit collision schemes*

By examining the results in Figs. 3, 4, and 6, it might seem counterintuitive that a larger transient error results from the SLIC+INT2 scheme in comparison to the SLIC+INT1 scheme. It could be assumed, after all, that second-order interpolation is inherently supposed to be more accurate than first-order interpolation. However, a more in-depth accuracy analysis can explain this phenomenon. To do so, it is necessary to re-examine the error accumulation of the transient solution of TGV flow. By checking Eqs. (37) and (38) for the analytical solution of the TGV flow, it can be seen that the decay of the flow is exclusively controlled by the term $e^{[-\nu(k_1^2+k_2^2)t]}$. Since $k_1$ and $k_2$ are constants, the actual decay is controlled by just two factors: the time $t$ and the viscosity $\nu$. Therefore, the longer the time and the larger the viscosity, the larger the decay would become. Translating this analysis to the numerical realm, this means that the error accumulation of the TGV transient solution is due to two (and only two) numerical ingredients: the time marching scheme that controls how time proceeds and the numerical viscosity that will alter the real viscosity of the flow. Therefore, the fact that SLIC+INT2 is less accurate than SLIC+INT1 for the temporal solution of the TGV flow, which is repeatedly shown in Figs. 3, 4, and 6, must be a manifestation of the combined effects of time marching (the implicit collision scheme is part of time marching) and numerical viscosity. As a result, it is necessary to see whether the SLIC+INT1 and SLIC+INT2 contribute the same amount of numerical viscosity prior to discussing their numerical error difference in temporal solutions observed in Figs. 3, 4, and 6.

In order to do this, a new flow case that can satisfy two conditions is needed. First, the flow must have a steady-state solution in order to remove the error difference due to time marching; and second, the solution of the flow must only be affected by viscosity from the physical point of view. To that end, the canonical lid-driven square cavity (LDSC) flow is chosen. By fixing the domain size and the velocity of the moving lid, the steady-state solution of the flow only depends on the viscosity. For example, the steady-state solution of Re=100 is shown by the solid line in Fig. 7. If the viscosity is decreased by 4, which will quadruple Re to 400 while keeping other parameters unchanged, the steady-state solution becomes what is shown as the dashed line. Therefore, it can be concluded that the flow profile will become more extreme when viscosity is decreased. By using this principle, the FVDBM steady-state solutions with all four implicit collision schemes on the Re=100 LDSC flow are obtained and compared in Fig. 8. From the enlarged insets in Fig. 8, it can be seen that the solutions with the TE, VT, and SLIC+INT1 schemes are identical, which indicates that these three schemes either produce zero or the same amount of numerical viscosity. On the other hand, the solution with the SLIC+INT2 scheme is slightly more disturbed than the other three approaches, which could only be the result of the SLIC+INT2 scheme producing slightly less numerical viscosity than SLIC+INT1 (and TE and VT), since all other parts of the model are kept the same. In fact, this is expected, since high-order spatial interpolation schemes will decrease numerical viscosity or diffusion error. By recalling the previous observations in Figs. 3, 4, and 6, we now can conclude that SLIC+INT2 introduces a much larger error to the time marching itself than SLIC+INT1. This error is so large that it offsets the benefit of the lower numerical viscosity compared to SLIC+INT1. Taken together, then, the SLIC+INT2 scheme produces larger errors than the SLIC+INT1 scheme for transient solutions. Therefore, for time marching only, SLIC+INT1 is superior to SLIC+INT2, even though SLIC+INT1 is mathematically much simpler. As for why the SLIC+INT1 is superior to SLIC+INT2 for the time marching itself, one is recommended to visit section 5.4 in this paper for a detailed explanation.

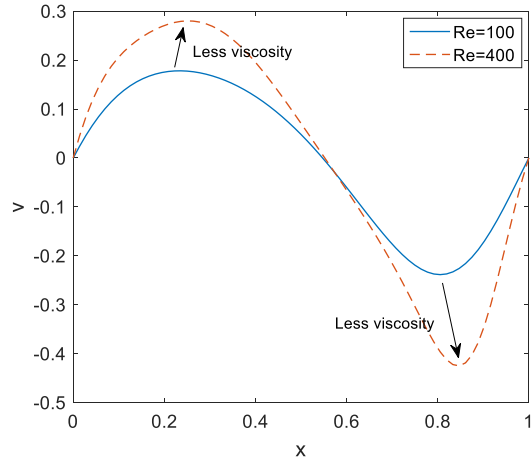

Figure 7. The effect of viscosity on lid-driven square cavity flow

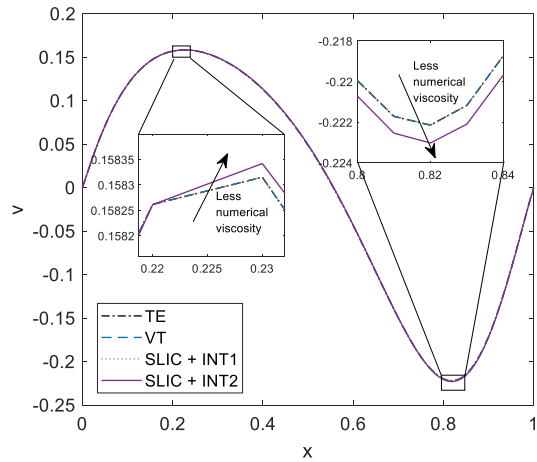

Figure 8. The numerical viscosities of different implicit collision schemes

**5.2 Computational cost**

The measured computational cost is the update time, $t_U$, in this paper, which is defined as:

$$t_U = \frac{t_T}{WS} \tag{43}$$

where $t_T$ is the total runtime for the simulation, $W$ is the total number of control volumes and $S$ is the total number of time steps or iterations. Therefore, $t_U$ is the overhead on all computational tasks in the FVDBM solver, not just the time spent on the implicit collision scheme. However, $t_U$ is able to reflect the difference in computational costs among different implicit collision schemes because all other numerical ingredients in the solver are the same. Table 1 lists the $t_U$ for the entire solver with different implicit collision schemes. All simulations are performed on an Intel i7-7700 3.6GHz CPU. Multiple measurements are taken and then averaged.

From the measurements, it can be seen that all of the schemes have almost the same $t_U$ except for the SLIC+INT2 approach. The SLIC+INT1 is slightly faster than the TE scheme because the former avoids the computational time needed for $f_\alpha^{eq,n+1}$ through Eq. (12), which is required in the TE scheme; the VT is slower than the TE scheme because the VT scheme requires two times of variable transformations, one of which is from *f* to *g* (Eq. (17)) and the other is from *g* to *f* (Eq. (19)). The SLIC+INT2 scheme is almost 25% slower than SLIC+INT1 due to the cumbersome second-order interpolation at multiple locations. The improvement in the computational cost of the SLIC+INT1 is comparatively mild, but that approach also comes with an accompanying gain in accuracy. Therefore, the SLIC+INT1 scheme can improve the accuracy with no extra computational cost.

*Table 1. Update time for the FVDBM solver with different implicit collision schemes*

| Implicit collision scheme | $t_U$ (Unit: second) |
|---|---|
| TE | $4.159 \times 10^{-5}$ |
| VT | $4.211 \times 10^{-5}$ |
| SLIC+INT1 | $4.145 \times 10^{-5}$ |
| SLIC+INT2 | $5.079 \times 10^{-5}$ |

**5.3 Stability**

As discussed in the Background section, the stability of solving the DBE is determined both by the advection and the collision. For the advection, the maximum $\Delta t$ is limited by $\Delta x$, which is the characteristic grid size and defined in the following equation for the triangular mesh used in this paper:

$$\Delta x = \sqrt{2V_{CV}} \qquad (44)$$

For the collision, the maximum $\Delta t$ is related to the relaxation time $\tau$. The stability region of each implicit collision scheme is shown in Fig. 9, in which ● and × represent stable and unstable points, respectively. First, by comparing Fig. 9(a) and 9(b), one can see that the VT scheme significantly improves the stability in the $\Delta t/\tau$ limit, which is changed from 2.6 to 100. A similar improvement was also reported in other work [35, 40]. Second, the comparison between Fig. 9(c)-(d) and 9(a) show that the SLIC schemes can also improve the $\Delta t/\tau$ limit by the same amount as the VT scheme. However, it is worth noting that the stability tests for the VT and SLIC schemes are purposely capped at $\Delta t/\tau = 100$ because that ratio is considered to be quite good in practice. Therefore, the VT and SLIC schemes may display different $\Delta t/\tau$ behaviors beyond 100, but that is not studied in the current paper. Finally, by comparing all of the sub-figures in Fig. 9, it is clear that all of the schemes share the same $\Delta t/\Delta x$ limit, which is 0.15. This is expected, since they produce the same amount of numerical viscosity (as shown in Fig. 8), with the exception that the SLIC+INT2 scheme generates a slightly lesser amount of numerical viscosity than the other schemes, which can be ignored eventually. It is well known that the stability for advection is closely related to the numerical viscosity. Therefore, these four schemes have an equal effect on the advection process and share the same $\Delta t/\Delta x$ limit based on the $\Delta t/\Delta x$ resolution used in the figure.

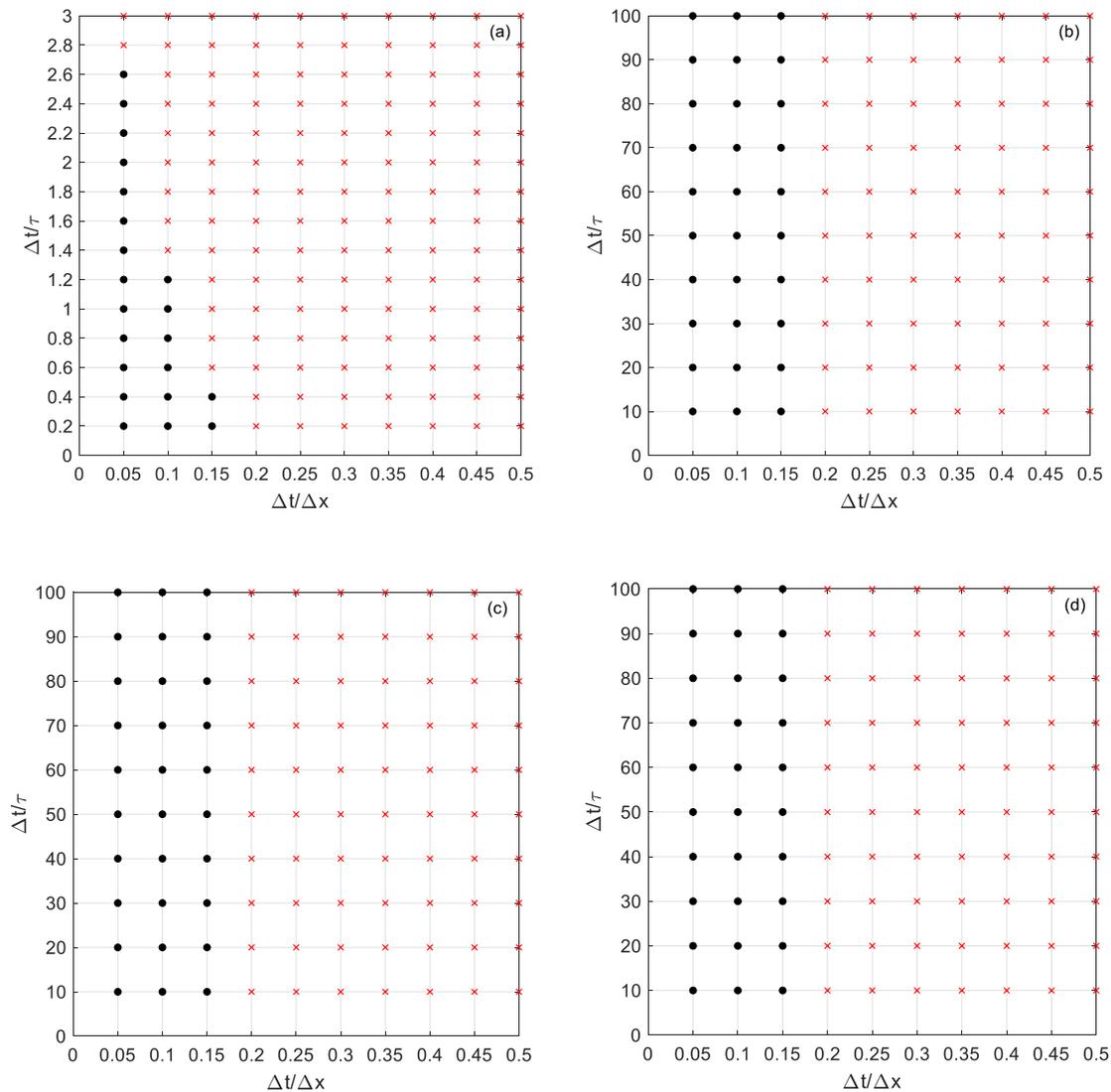

Figure 9. The stability regions of the FVDBM solutions with different implicit collision schemes for (a) TE; (b) VT; (c) SLIC+INT1; and (d) SLIC+INT2

**5.4 Preliminary conclusions and further discussions from a different perspective**

Some preliminary conclusions can be made based on the results gathered so far. The comparison in accuracy reveals that, first, the VT scheme is slightly more accurate than the TE scheme, while both satisfy Eq. (41). Second, the SLIC+INT1 scheme is much more accurate than the TE and VT schemes, and satisfies Eq. (42). In the tested range of $\Delta t$, it can improve the accuracy by a factor of eight at maximum and by a factor of four on average. Third, the SLIC+INT2 scheme satisfies Eq. (42) as well, and is also an improvement over the TE and VT schemes, but is not as accurate as SLIC+INT1. The deterioration is exclusively due to the fact that the SLIC+INT2 approach generates a higher level of error during time marching (although it produces slightly less numerical viscosity). The tests on computational costs show that, first, the SLIC+INT1 is the fastest scheme because it is mathematically the simplest; second, upgrading the interpolation from first order to second order for the SLIC scheme is not a good investment given the fact it unpleasantly cuts the accuracy. At last, the stability test shows that the two SLIC schemes have the same $\Delta t/\tau$ limit as the VT scheme in the tested range of $\Delta t/\tau$, which is an order of magnitude higher than the TE scheme. In conclusion, SLIC+INT1 is the most advantageous scheme to use, considering the accuracy, computational cost, and stability all at the same time.

By comparing the mathematical forms of the TE scheme (Eq. (12)) and the SLIC+INT1 scheme (Eq. (32)) side by side, one may draw an erroneous conclusion that these two schemes belong to the same mathematical family based on the temporal extrapolation of equilibrium PDFs. This is because Eq. (12) is a second-order extrapolation that utilizes the equilibrium PDFs at the two previous time steps $t_n$ and $t_{n-1}$, and Eq. (32) appears to be a first-order temporal extrapolation that uses the equilibrium PDFs at only one time step, $t_n$. However, Eq. (32) actually is not connected to a temporal extrapolation at all. There are two supporting pieces of evidence. First, if the SLIC+INT1 and TE schemes belong to the same family but have different orders of accuracy, one would have been able to see that the SLIC+INT1 generates a higher, not smaller, error than the TE scheme in Fig. 4. Second, the vigorous derivation in Section 4 reveals the true origin of Eq. (32), which is completely unrelated to temporal extrapolation. The reason why it appears to be a first-order temporal extrapolation scheme is only because Eq. (32) is a special case that applies the first-order spatial interpolation to the general form of the semi-Lagrangian BGK model (Eq. (29)).

In order to explain why the SLIC+INT1 scheme, which bears the simplest mathematical form, can outperform other schemes (especially the SLIC+INT2 approach) in terms of the accuracy for time marching itself, it is necessary to re-examine all of the schemes presented in this paper from a different perspective. The four schemes - TE, VT, SLIC+INT1, and SLIC+INT2 - can be re-formulated and generalized such that:

$$f_\alpha^{eq,n+1} = f_\alpha^{eq,n} + \delta \quad (45)$$

where

$$\delta = \begin{cases} f_\alpha^{eq,n} - f_\alpha^{eq,n-1}, & \text{for TE (please refer to Eq. (12))} \\ -M\{m[F_\alpha^n]\}\Delta t, & \text{for VT (please refer to Eqs. (18), (21), and (23))} \\ 0, & \text{for SLIC + INT1 (please refer to Eq. (32))} \\ \text{a nontrivial term related to Eq. (33)}, & \text{for SLIC + INT2} \end{cases} \quad (46)$$

It can be easily concluded that $\delta$ is a non-zero term for all schemes except for the SLIC+INT1 approach. Therefore, we can propose the following hypothesis:

*The temporal error is related to the size of $\delta$ and is minimized when $\delta = 0$* (H1)

If H1 is true, the reason why the SLIC+INT1 scheme exhibits the best performance is self-explanatory. However, for H1 to be valid, another hypothesis first must be valid: on the system level, $\delta = 0$ represents the absolutely true case. In other words:

*If a simulation is correct, globally, $f_\alpha^{eq,n+1} - f_\alpha^{eq,n} \to 0$ as time progresses, or $\frac{\partial f_\alpha^{eq}}{\partial t} \to 0$ as $t \to \infty$* (H2)

In order to prove H2 and then H1, we can develop the following numerical evidence. A standard LBM simulation is performed on the same TGV flow. It is used as a benchmark, since its computational paradigm is completely different from the FVDBM and it does not require the computation of the implicit $f_\alpha^{eq,n+1}$. The $\frac{\partial f_\alpha^{eq}}{\partial t}$ at any location $x$ and time step $t_n$ is computed numerically as $\frac{f_\alpha^{eq,n}(x) - f_\alpha^{eq,n-1}(x)}{\Delta t}$ at each grid point, and then averaged over the entire computational domain. The calculation is performed on each of the nine total directions, and then plotted over time. The results from the LBM simulation as well as the FVDBM simulations with all of the studied implicit schemes are shown in Fig. 10.

It can be seen from the results that there are three pieces of evidence to support H2. First, the values of $\frac{\partial f_\alpha^{eq}}{\partial t}$ for all of the numerical schemes (LBM as well as FVDBM) on all nine directions are very small. Apart from direction 0, in which the value is of order $10^{-9}$, the values of $\frac{\partial f_\alpha^{eq}}{\partial t}$ in the other eight directions are on the order of $10^{-10}$. Second, for the LBM, $\frac{\partial f_\alpha^{eq}}{\partial t}$ is oscillating, but the magnitude of the oscillation is decreasing over time in all directions; and for the FVDBM, the results from all implicit schemes are monotonically decreasing over time except at the beginning of the simulations (the simulations are initialized with $f_\alpha^{eq}$, and this is why $\frac{\partial f_\alpha^{eq}}{\partial t} = 0$ at $t = 0$). Third, for the LBM simulation, the centerlines of the oscillations of $\frac{\partial f_\alpha^{eq}}{\partial t}$ are computed, and are horizontal lines passing through trivial values with very small magnitudes. In direction 0, this value is $4.02 \times 10^{-11}$; and it is $-4.6 \times 10^{-12}$ for directions 1 to 4 and $-3.9 \times 10^{-12}$ and directions 5 to 8, respectively. These values are near zero, which is another indicator that $\frac{\partial f_\alpha^{eq}}{\partial t} \to 0$ on the global scale captures the true nature of the simulation during time marching.

Given that H2 is true, the validity of H1 can be proven with ease. It can be seen that in all nine directions, the SLIC+INT1 scheme consistently generates a smaller size for $\frac{\partial f_\alpha^{eq}}{\partial t}$ than the other three implicit schemes. And at the same time, the temporal error of the SLIC+INT1 scheme for time marching is the smallest, which is already shown in Fig. 4. So, H1 is true as well.

The theoretical reason why $\frac{\partial f_\alpha^{eq}}{\partial t} \to 0$ on the global scale (H2) is also provided here, which is embedded in the DBE itself. By using the material derivative $D/Dt = \partial/\partial t + \boldsymbol{e} \cdot \boldsymbol{\nabla}$, the DBE with the BGK collision model (Eq. (1)) can be rewritten as:

$$\frac{Df_\alpha}{Dt} = -\frac{1}{\tau}\left(f_\alpha - f_\alpha^{eq}\right) \tag{47}$$

Since it always holds that $f_\alpha = f_\alpha^{eq} + f_\alpha^{neq}$, Eq. (47) then becomes:

$$\frac{Df_\alpha^{eq}}{Dt} = -\left(\frac{Df_\alpha^{neq}}{Dt} + \frac{1}{\tau}f_\alpha^{neq}\right) \tag{48}$$

There are two components in the parenthesis on the RHS of Eq. (48). The first one is $\frac{Df_\alpha^{neq}}{Dt}$, which is controlled by streaming; and the second part is $\frac{1}{\tau}f_\alpha^{neq}$, which is governed by the relaxation process. These two components tend to cancel out each other as a result of the famous "ever-shifting battle" originally evoked by Ludwig Boltzmann. The explanation for what this means and why it happens can be found in [7]. The purpose of relaxation is to achieve a balance in which the distribution functions are couched into the universal local Maxwell-Boltzmann distribution. On the other hand, streaming operates in a completely opposite way. It destroys the balance established by the relaxation by reviving non-equilibrium through inhomogeneity. As a result, the sum of $\frac{Df_\alpha^{neq}}{Dt}$ and $\frac{1}{\tau}f_\alpha^{neq}$ will converge to zero as time progresses. However, when the numerical viscosity is zero (such as in the LBM), the RHS of Eq. (48) will never monotonically converge to zero, but always oscillates within a narrowing band (capped by very small values) due to the "ever-shifting" nature of the balance, as seen in the oscillatory curves of Fig. 10. Therefore, $\frac{Df_\alpha^{eq}}{Dt}$, which is the left-hand-side (LHS) of Eq. (48), will inherit the same converging nature exhibited by the RHS as a result of LHS=RHS (the minus sign on the RHS does not affect this nature; and the LHS is also oscillatory if numerical viscosity is zero). When $\frac{Df_\alpha^{eq}}{Dt} \to 0$ locally, it means that $f_\alpha^{eq}$ only slightly departs from exact conservation along the light cone (the lattice directions or characteristics) as time passes. On the local scale, $\frac{Df_\alpha^{eq}}{Dt} \to 0$ and $\frac{\partial f_\alpha^{eq}}{\partial t} \to 0$ are not the same, since the former is in Lagrangian space while the latter is in Eulerian space. However, on a global scale, in which a global average is taken, $\frac{Df_\alpha^{eq}}{Dt} \to 0$ can be considered equivalent to $\frac{\partial f_\alpha^{eq}}{\partial t} \to 0$, further validating H2 on a theoretical basis, beyond the numerical evidence shown in Fig. 10.

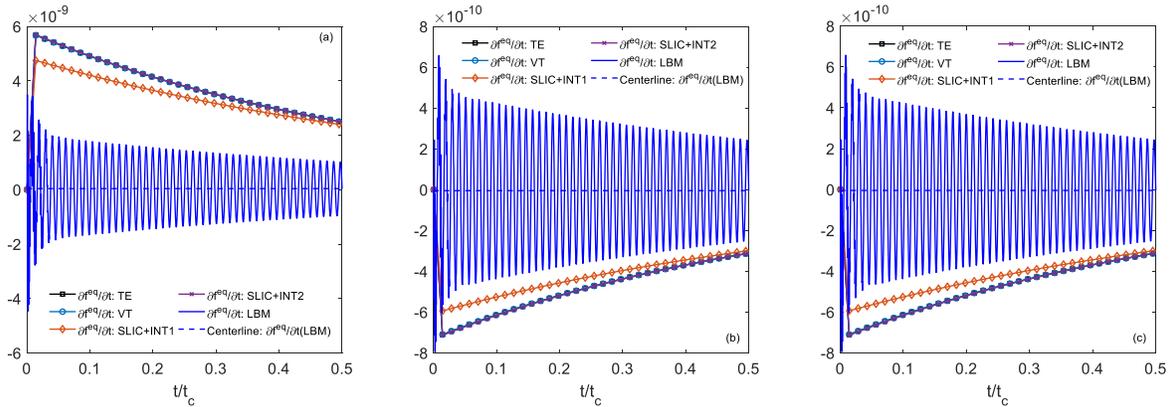

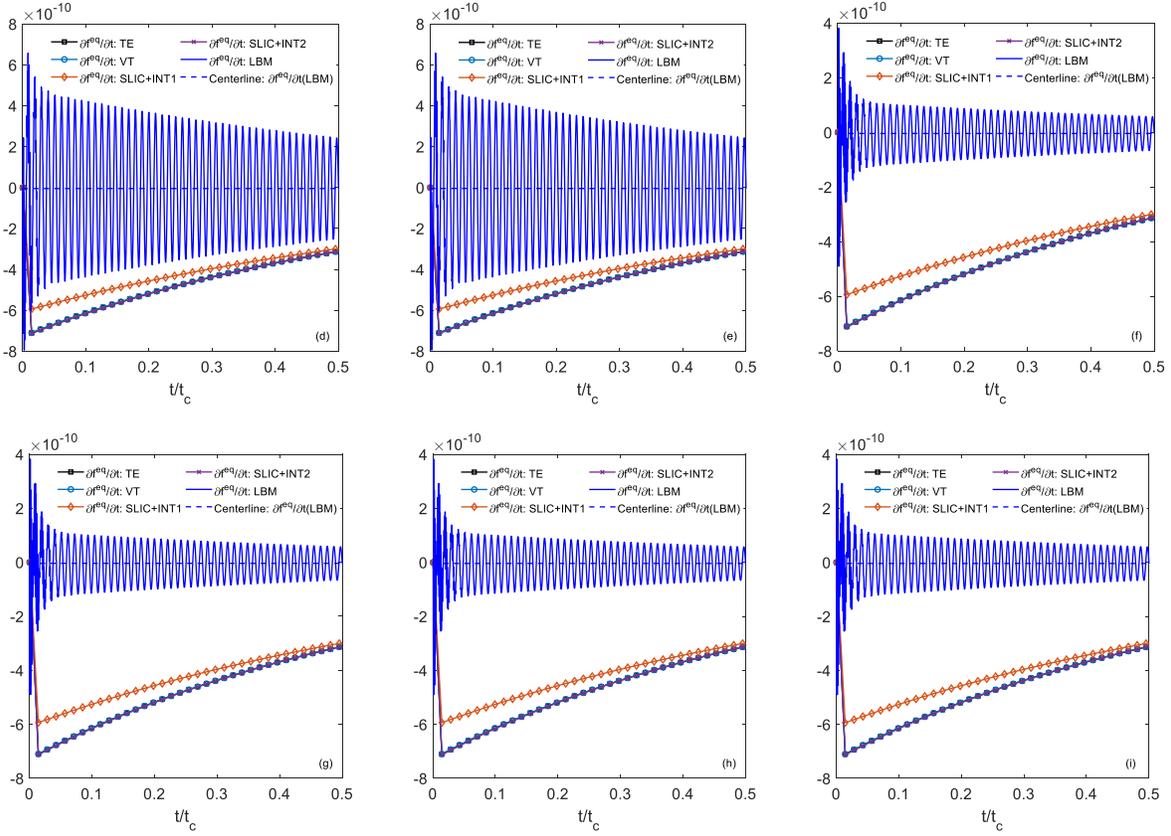

*Figure 10. The transient behaviors of $\partial f_\alpha^{eq}/\partial t$ for different models on the Taylor-Green vortex flow for (a) direction 0; (b) direction 1; (c) direction 2; (d) direction 3; (e) direction 4; (f) direction 5; (g) direction 6; (h) direction 7; (i) direction 8*

## 6. The application of the SLIC+INT1 model: a secondary study

In terms of implementing the SLIC+INT1 scheme for different simulations, there is a simple rule of thumb to do this:

*Replace $f_\alpha^{eq,n+1}$ with $f_\alpha^{eq,n}$ in any place it appears.*

It is important to recall that the numerical results and conclusions for the implicit collision schemes in the previous section are made in a specific numerical context that the time marching of the FVDBM is chosen to have $\theta = 1$ in Eq. (9) and the flux term $F_\alpha^n$ is completed with the second-order upwind scheme (SOU). In order to demonstrate that the numerical advantages of the SLIC+INT1 scheme in terms of accuracy, computational cost and stability will generally hold, the same comparisons from the previous section will be performed again, but this time on the FVDBM with a different time marching scheme and a different flux calculation. Starting from Eq. (9) again for the general FVDBM, and choosing $\theta = \frac{1}{2}$, which is a popular choice for a second-order accuracy [22, 23, 35-40], Eq. (9) then becomes:

$$T_\alpha^n = \frac{1}{2}(C_\alpha^n + C_\alpha^{n+1}) - F_\alpha^n \qquad (49)$$

After using Eq. (4) to expand the BGK collision terms $C_\alpha^n$ and $C_\alpha^{n+1}$ and using Eq. (7) to replace $T_\alpha^n$, Eq. (49) is further transformed to:

$$f_\alpha^{n+1} = f_\alpha^n - \frac{\Delta t}{2\tau}\left[\left(f_\alpha^n - f_\alpha^{eq,n}\right) + \left(f_\alpha^{n+1} - f_\alpha^{eq,n+1}\right)\right] - \Delta t F_\alpha^n \tag{50}$$

After combining the terms that contain $f_\alpha^n$ and $f_\alpha^{n+1}$, Eq. (50) becomes:

$$f_\alpha^{n+1} = \frac{2\tau - \Delta t}{2\tau + \Delta t} f_\alpha^n + \frac{\Delta t}{2\tau + \Delta t}\left(f_\alpha^{eq,n} + f_\alpha^{eq,n+1}\right) - \frac{2\tau \Delta t}{2\tau + \Delta t} F_\alpha^n \tag{51}$$

Now, instead of using the SOU for $F_\alpha^n$, it is calculated with the piece-wise linear (PL) Godunov-type flux scheme (which is also second-order) developed in [27]. The PL flux scheme produces much less diffusion error than the SOU approach because it calculates the averaged flux from $t_n$ to $t_{n+1}$. As a result, the PL flux scheme is also a function of $\Delta t$, so that the flux calculation will become more accurate as $\Delta t$ becomes smaller (please refer to [27] for more details). It can be seen that there is still unresolved implicitness embedded in $f_\alpha^{eq,n+1}$ in Eq. (51) (and Eq. (50)), which will be treated with different schemes in the following subsections.

**6.1 The temporal extrapolation (TE) scheme**

The same numerical procedure in Section 3.1 can be repeated here:
Step 1: Calculate the moments with Eq. (14) with the newest $f_\alpha$;
Step 2: Calculate $f_\alpha^{eq}$ with Eq. (13) based on the moments from step 1;
Step 3: Calculate $f_\alpha^{eq,n+1}$ with Eq. (12);
Step 4: Update $f_\alpha$ with Eq. (51).

**6.2 The variable transformation (VT) scheme**

Since a different $\theta$ value is used, the variable transformation process should be redone accordingly from the beginning. By defining another new variable $h_\alpha$ as:

$$h_\alpha = f_\alpha - \frac{\Delta t}{2} C_\alpha \tag{52}$$

we have:

$$h_\alpha^{n+1} = f_\alpha^{n+1} - \frac{\Delta t}{2} C_\alpha^{n+1} = f_\alpha^{n+1} + \frac{\Delta t}{2\tau}\left(f_\alpha^{n+1} - f_\alpha^{eq,n+1}\right) \tag{53}$$

Eq. (50) can then be rewritten as:

$$h_\alpha^{n+1} = \frac{2\tau - \Delta t}{2\tau} f_\alpha^n + \frac{\Delta t}{2\tau} f_\alpha^{eq,n} - \Delta t F_\alpha^n \tag{54}$$

After $h_\alpha^{n+1}$ is computed, $f_\alpha^{n+1}$ can be recovered by using Eq. (53) such that:

$$f_\alpha^{n+1} = \frac{2\tau}{2\tau + \Delta t}\left(h_\alpha^{n+1} + \frac{\Delta t}{2\tau} f_\alpha^{eq,n+1}\right) \tag{55}$$

Again, $f_\alpha^{eq,n+1}$ can be resolved by the following relation because the new variable $h$ has the same moments as $f$:

$$f_\alpha^{eq,n+1} = h_\alpha^{eq,n+1} \tag{56}$$

Then $h_\alpha^{eq,n+1}$ is computed by Eq. (13) with the moments that are calculated by:

$$\begin{bmatrix}\rho\\\rho\bm{u}\end{bmatrix} = \Sigma_{\alpha=0}^{N-1}\begin{bmatrix}h_\alpha\\\bm{e}_\alpha h_\alpha\end{bmatrix} \tag{57}$$

As a result, the procedure of using the VT scheme on the new FVDBM is:

Step 1: Calculate the moments with Eq. (14) with the newest $f_\alpha$ ;
Step 2: Calculate $f_\alpha^{eq,n}$ with Eq. (13) based on the moments from step 1;
Step 3: Calculate $h_\alpha^{n+1}$ with Eq. (54);
Step 4: Calculate the moments with Eq. (57) based on the $h_\alpha^{n+1}$ obtained from step 3;
Step 5: Calculate $h_\alpha^{eq,n+1}$ with Eq. (13) based on the moments from step 4;
Step 6: Update $f_\alpha$ with Eq. (55) by applying Eq. (56).

By comparing the numerical procedures of the new and the old FVDBM in Section 3.2, it can be seen that the new one requires the calculation of $f_\alpha^{eq,n}$, which does not appear in the older approach. This is solely because both $f_\alpha^{eq,n}$ and $f_\alpha^{eq,n+1}$ are needed to update $f_\alpha$ in the new FVDBM in which $\theta = \frac{1}{2}$.

### 6.3. The semi-Lagrangian implicit collision (SLIC) schemes

The calculation of the new FVDBM with the SLIC+INT1 scheme becomes very simple:
Step 1: Calculate the moments with Eq. (14) with the latest $f_\alpha$ ;
Step 2: Calculate $f_\alpha^{eq,n}$ with Eq. (13) based on the moments from step 1;
Step 3: Update $f_\alpha$ with Eq. (51) using the rule of thumb, which can further reduce Eq. (51) to:

$$f_\alpha^{n+1} = \frac{2\tau-\Delta t}{2\tau+\Delta t}f_\alpha^n + \frac{2\Delta t}{2\tau+\Delta t}f_\alpha^{eq,n} - \frac{2\tau\Delta t}{2\tau+\Delta t}F_\alpha^n \tag{58}$$

In order to explore a comparison between the SLIC+INT1 and SLIC+INT2 schemes for this new numerical case, the calculation with the SLIC+INT2 is also provided, utilizing the following steps:

Step 1: Calculate the coefficients with Eq. (34);
Step 2: Calculate the PDF at the tracked-back location with Eq. (33);
Step 3: Repeat step 1 and step 2 for all tracked-back locations;
Step 4: Gather the PDFs at all tracked-backed locations and plug them into Eq. (29) to compute $f_\alpha^{eq,n+1}$;
Step 5: Update $f_\alpha$ with Eq. (51) by using the $f_\alpha^{eq,n+1}$ from step 4.

### 6.4 The VT scheme modified by SLIC+INT1

As stated in the rule of thumb, $f_\alpha^{eq,n+1}$ can be replaced by $f_\alpha^{eq,n}$ anywhere it appears, including in Eq. (55) for the VT scheme. As a result, the calculations of the moments based on $h$ (Eq. (57)), the subsequent calculation of $h_\alpha^{eq,n+1}$ (Eq. (13)), and the final update of $f_\alpha^{eq,n+1}$ (Eq. (56)) can be completely avoided. Therefore, the VT scheme modified by the SLIC+INT1 scheme can be realized in the following sequence:

Step 1: Calculate the moments with Eq. (14) with the newest $f_\alpha$ ;
Step 2: Calculate $f_\alpha^{eq,n}$ with Eq. (13) based on the moments from step 1;
Step 3: Calculate $h_\alpha^{n+1}$ with Eq. (54);
Step 4: Update $f_\alpha$ with Eq. (55) by using the rule of thumb.

However, it is worth noting that the modified VT scheme will generate exactly the same solution as the SLIC+INT1 scheme, because after replacing $h_\alpha^{n+1}$ with Eq. (54), Eq. (55) becomes Eq. (51) automatically. After that, substituting $f_\alpha^{eq,n+1}$ with $f_\alpha^{eq,n}$ will have the same effect as the SLIC+INT1 scheme.

### 6.5 Numerical comparisons and discussions

The five schemes, TE, VT, SLIC+INT1, SLIC+INT2, and VT modified by SLIC+INT1 will be compared in this subsection in terms of accuracy, computational cost, and stability. The numerical settings for the new comparison are kept the same as in Section 5.

### 6.5.1 Accuracy

The errors of transient solution from 0 to $0.5t_c$ of the TGV flow for the five schemes on the new FVDBM (new time marching and new flux scheme) with four sizes of $\Delta t$ are compared in Fig. 11. By comparing Fig. 11 with Fig. 4, several quick observations can be made. First, all errors shown in Fig. 11 are smaller than those in Fig. 4. This is because the PL flux scheme can produce much less numerical viscosity than the SOU flux. As discussed at the end of Section 5.1, numerical viscosity is one of the two factors that can influence the accuracy of the transient solution of TGV flow. A smaller numerical viscosity will make the transient solution more accurate. Second, TE and VT schemes still produce the same results. Third, the SLIC+INT1 is still the most accurate scheme. In the tested range of $\Delta t$, the maximum factor of accuracy improvement is more than five at $\Delta t = 0.2\tau$, compared to the TE or VT scheme. On average, it can still improve the accuracy by a factor of four. Fourth, the SLIC+INT2 is still not as accurate as the SLIC+INT1 approach for the same reason (Eq. (46)). In addition, in Fig. 11(a) in which $\Delta t$ is small, it can be seen that even the TE and VT schemes are better choices than the SLIC+INT2 approach.

There are also some other new phenomena. First, after being modified by the SLIC+INT1, VT can also improve its accuracy. From the results, it can be seen that the SLIC+INT1 and the modified VT produce the same accuracy, which echoes the analysis at the end of Section 6.4. Second, all schemes in Fig. 11 have an increasing error with an increase in $\Delta t$, which differs from the behavior observed in Fig. 4. This can be more easily seen in Fig. 12, in which the percentage errors of different schemes are plotted against different sizes of $\Delta t$. The upward trend is due to the fact that the error of the PL Godunov flux scheme increases with an increase in $\Delta t$, which changes the nature of the entire FVDBM solver to the one that satisfies the following condition, no matter which implicit collision scheme is used:

$$\frac{\epsilon_1}{\epsilon_2} < \frac{\Delta t_1}{\Delta t_2} \qquad (59)$$

However, it is interesting to note that in Fig. 12 the errors of the modified VT and two SLIC schemes grow at a slower rate than the other schemes when $\Delta t$ is increasing. This is because although the PL Godunov flux scheme generates a higher error with a larger $\Delta t$, the SLIC and the modified VT implicit collision schemes still decrease the error with a growing $\Delta t$ (Eq. (42)) while the TE and VT are still not affected by $\Delta t$ (Eq. (41)). As a result, the combined effects of the flux scheme and the implicit collision scheme will behave as what is shown in Fig. 12.

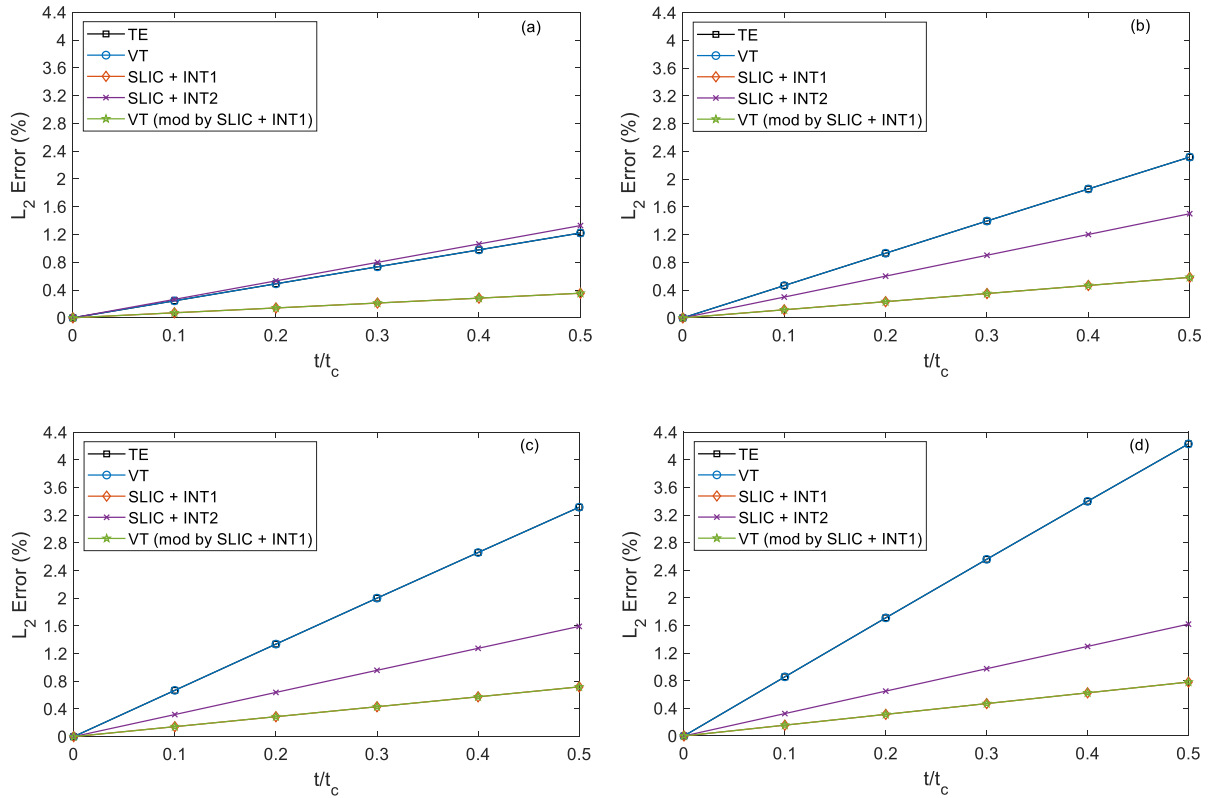

Figure 11. The $L_2$ errors of the transient solutions of the new FVDBM with different implicit collision schemes during the time span from 0 to $0.5t_c$ for (a) $\Delta t = 0.05\tau$; (b) $\Delta t = 0.1\tau$; (c) $\Delta t = 0.15\tau$; and (d) $\Delta t = 0.2\tau$

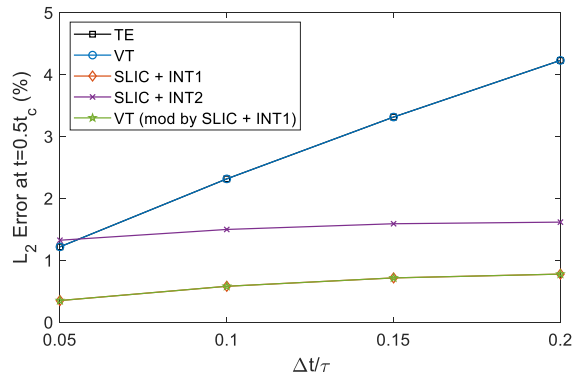

*Figure 12. The effects of $\Delta t$ on the errors of the new FVDBM solutions with different implicit collision schemes*

### 6.5.2 Computational cost

The update time $t_U$ (Eq. (43)) of the new FVDBM solver with each of the five implicit schemes is measured and listed in Table 2. Again, it can be seen that the SLIC+INT1 scheme is still the fastest. In addition, the modified VT scheme is faster than the original VT scheme due to the saved computations, but still slower than the SLIC+INT1. However, there are two major differences compared to the observations in Table 1 for the previous FVDBM solutions. First, the update times for the new FVDBM solver with all implicit collision models are universally shorter than their counterparts in Table 1. This is because the PL Godunov flux scheme is faster than the SOU flux scheme. Second, in Table 1, the SLIC+INT1 is roughly 2% faster than the VT scheme, but for the new FVDBM, the SLIC+INT1 presents a 13% improvement in speed. The reason for this is that in the old FVDVM, in which $\theta = 1$, both the VT and SLIC+INT1 schemes require just one calculation of the equilibrium PDF. However, the VT scheme in the new FVDBM with $\theta = \frac{1}{2}$ requires the calculation of the equilibrium PDF two times, which was also stated at the end of Section 6.2, while the SLIC+INT1 scheme still just requires one calculation. Since the equilibrium PDF calculation is very costly, the difference in speed becomes larger. Finally, the SLIC+INT2 scheme is still the slowest due to its much larger computational load.

*Table 2. Update time for the new FVDBM solver with different implicit collision schemes*

| Implicit collision scheme | $t_U$ (Unit: second) |
|---|---|
| TE | $3.589 \times 10^{-5}$ |
| VT | $4.064 \times 10^{-5}$ |
| SLIC+INT1 | $3.521 \times 10^{-5}$ |
| SLIC+INT2 | $4.426 \times 10^{-5}$ |
| VT modified by SLIC+INT1 | $3.762 \times 10^{-5}$ |

### 6.5.3 Stability

The last test is the stability test on the new FVDBM with the five implicit collision schemes. The stability region for each scheme is shown in Fig. 13. The first difference from the results shown in Fig. 9 is that the $\Delta t/\Delta x$ limit becomes larger. This is the advantage of the PL Godunov flux scheme over the SOU flux scheme. Consistent with Fig. 9, it can be seen as well that the two SLIC schemes have a much higher $\Delta t/\tau$ limit than the TE scheme and have the same $\Delta t/\tau$ limit as the VT scheme. Additionally, the stability of the modified VT does not change compared to its original version, and also stays the same as the SLIC+INT1 since the SLIC+INT1 and modified VT scheme are mathematically the same, as explained as the end of Section 6.4.

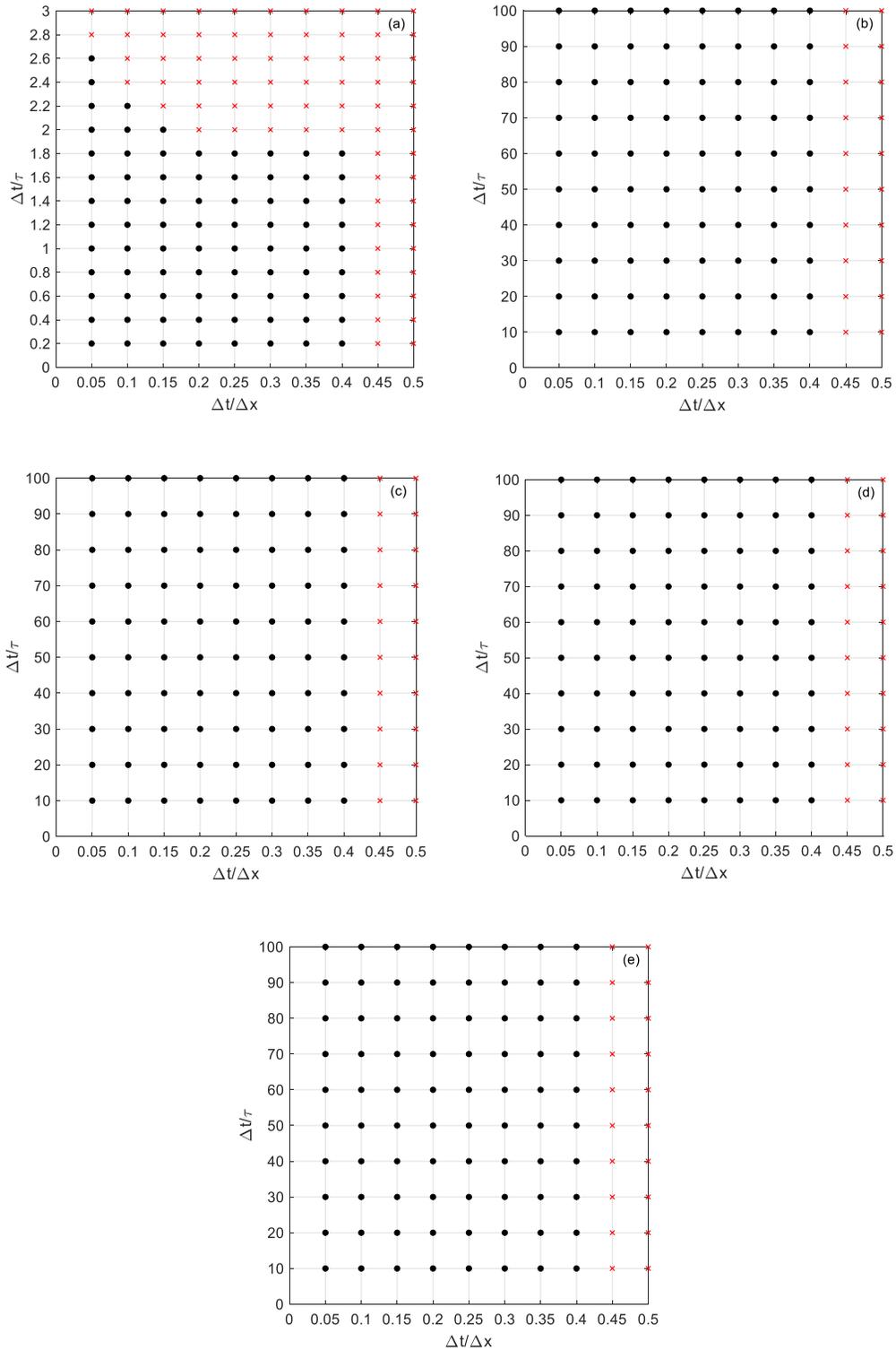

*Figure 13. The stability regions of the new FVDBM solutions with different implicit collision schemes for (a) TE; (b) VT; (c) SLIC+INT1; (d) SLIC+INT2; and (e) VT modified by SLIC+INT1*

**7. Conclusions**

In this paper, a new method to resolve the implicit BGK collision is developed for the finite volume discrete Boltzmann method (FVDBM). This new method stems from applying the semi-Lagrangian approach to the implicit equilibrium PDFs in the BGK collision. With the help of the first-order interpolation, the new scheme becomes as simple as enforcing $f_\alpha^{eq,n+1} = f_\alpha^{eq,n}$. By comparing this new scheme with two existing ones, the temporal extrapolation (TE) and the variable transformation (VT) approaches, on the FVDBM with different time marching and flux calculations schemes, three advantages consistently can be demonstrated:

1)  The new scheme can improve the temporal accuracy by almost an order of magnitude in the tested range of numerical settings while not affecting the spatial accuracy;

2)  The new scheme can slightly lower the computational cost (so that it can be concluded that the new scheme can improve accuracy at no extra cost); and

3) The new scheme can significantly improve the stability in the $\Delta t/\tau$ limit compared to the TE scheme, and maintain the same $\Delta t/\tau$ limit as the VT scheme. In addition, the new scheme does not affect the $\Delta t/\Delta x$ limit.

In addition, this paper also demonstrates that the new approach can be easily applied in any place where $f_\alpha^{eq,n+1}$ is present by modifying the VT scheme with it. The test results show that the modified VT scheme still presents the three advantages listed above.

In order to test whether the accuracy could be improved further, a second-order interpolation scheme was also developed and applied. However, a detailed analysis shows that using second-order interpolation unexpectedly displays no advantage over the simple first-order interpolation approach. It is found that the accuracy is decreased while the computational cost is increased.

In order to explain why the semi-Lagrangian approach with the simple first-order interpolation (SLIC+INT1) outperforms other schemes, especially the same semi-Lagrangian scheme but with the second-order interpolation (SLIC+INT2), two hypotheses have been successfully validated by numerical evidence as well as a theoretical analysis. The detailed analysis showed that $f_\alpha^{eq,n+1} - f_\alpha^{eq,n}$ must converge to zero during time marching for a correct simulation. The SLIC+INT1 scheme is the only method that introduces the smallest value of $f_\alpha^{eq,n+1} - f_\alpha^{eq,n}$, which is why it is the most accurate approach. Therefore, the rule of converging $f_\alpha^{eq,n+1} - f_\alpha^{eq,n}$ can also be used as a guideline to develop new time marching schemes for future DBM-based methodologies.

## 8. Acknowledgements

Special thanks to Dr. Jing-Mei Qiu from the University of Delaware for constructive discussion and advice. The authors also thank the National Science Foundation for supporting this work under Grant No. CBET-1233106, as well as acknowledging funding from the European Research Council under the Horizon 2020 Program Grant Agreement n. 739964 ("COPMAT").